\documentclass[doublecol,linenumbers]{epl2}
\usepackage{graphicx}
\usepackage[english]{babel}
\usepackage{latexsym,amsmath,amscd,amssymb,graphicx,amsbsy,color,xcolor,amsfonts,amsthm}
\title{ Relativistic  Newtonian Dynamics under a Central Force}
\shorttitle{Relativistic  Newtonian Dynamics}
\author{Yaakov Friedman}
\institute{Jerusalem College of
Technology, P.O.B. 16031 Jerusalem 91160, Israel}

\abstract{Planck's formula and General Relativity indicate that potential energy influences spacetime. Using Einstein's Equivalence Principle and an extension of his Chock Hypothesis, an explicit description of this influence is derived.  We present a new relativity model by incorporating the influence of the potential energy on spacetime in  Newton's dynamics for  motion under a central force.  This model extends the model used by Friedman and Steiner \cite{FS} to obtain the exact precession of Mercury without curving spacetime.  We also present a solution of this model for a hydrogen-like atom, which explains the reason for a probabilistic description.}

\begin{document}
\maketitle

\section{Introduction}

 Newtonian dynamics describes properly the motion of objects in absolute space and time \cite{GPS},\cite{McCuskey}. Relativistic dynamics can be considered as a modification of  Newtonian dynamics  {to describe the motion in} influenced spacetime \cite{Rindler}.  Special Relativity (SR) dynamics \cite{E05} time and space are influenced {by} velocity or equivalently by the kinetic energy. Is the potential energy also influencing spacetime and dynamic laws?

 Planck's formula indicates that time is influenced by any type of energy. In General Relativity (GR), space and time are influenced by the gravitational potential. As we will show later the  anomalous precession of the perihelion of Mercury implies that space is also influenced by gravitational energy. What about the influence on spacetime of non-gravitational potentials?

 There are several theories predicting universal \cite{Sakharov},\cite{FG10} or system dependent \cite{CAIANIELLO} maximal accelerations and the influence of acceleration or potential energy on spacetime \cite{SCADPETTA},\cite{Torrome}. These theories predict \cite{F_Ann} a new relativistic dynamics which has interesting predictions for harmonic oscillations \cite{F_HO} and for a hydrogen-like atom \cite{FR}. Recently the dynamic equation for planetary motion using the idea of the influence of the gravitational potential on spacetime which led to prediction of the accurate anomalous precession of  Mercury without the need of curving spacetime was obtained in \cite{FS}. However,  the influence on  time expressed by the time dilation factor, differs from the one obtained in the maximal acceleration models. The recent YARK model (\cite{Yarman} and references therein) also describes the influence of potential energy on spacetime. But unlike in \cite{FS}, where this influence on space differs in different directions (the cause of the precession), in YARK this influence, expressed by the metric (formula (17)), acts the same way on all spacial directions.

 In this paper we extend  the Relativistic Newtonian Dynamics (RND) introduced in \cite{FS} for planetary motion. We present the Relativistic Newtonian Dynamics for particle motion under a conservative central force in flat spacetime by incorporating the influence of the potential (not only the gravitational one) energy on spacetime. This influence will be derived from the known principles introduced by A. Einstein, but will not need curving the spacetime.

\section{Newtonian dynamics under conservative force}

We will recall the description of motion of a mass particle under a conservative force in Newtonian dynamics. Pick an inertial lab frame $K$ with coordinates $\mathbf{x}=(x^1,x^2,x^3)$ and time $t$. We denote the mass of the particle $m$, which will be assumed to be constant (rest mass), and denote the force $\mathbf{F}(\mathbf{x})$ and its potential $U(\mathbf{x}).$   The trajectory $\mathbf{x}(t)$ of the  motion of the particle is described by the Newton's second law
\begin{equation}\label{2Newton}
 m\frac{d^2\mathbf{x}(t)}{dt^2}=\mathbf{F}(\mathbf{x}(t))=-\nabla U(\mathbf{x}(t)),
\end{equation}
where  $\nabla U$ denotes the gradient of $U$.

From this equation it follows \cite{GPS} that the total energy $E$, which is a function of the state, defined by
\begin{equation}\label{energycos}
  E(\mathbf{x},\mathbf{v})=\frac{m}{2}\mathbf{v}^2+ U(\mathbf{x})
\end{equation}
is conserved during the motion. The first term on the right-hand side is the kinetic energy and the second one is the potential energy.
From (\ref{2Newton}), the potential $U$ is defined up to a constant. We will need a uniquely defined $U$ with the condition $U(O)=0$ where the point $O$ is defined by $\nabla U (O)=0$. For planetary motion or any inverse square force potential, $O$ is at $\infty$ and for a harmonic oscillator it is at the origin of the oscillator.

For any energy associated with a mass particle  we define its  unit-free \textit{dimensionless energy}  as the ratio of this energy to  $\frac{1}{2}mc^2$, where $c$ the speed of light. The  \textit{dimensionless kinetic energy} (DKE) is $\frac{1}{c^2}\left(\frac{d\mathbf{x}}{dt}\right)^2=\beta^2$, where $\boldsymbol{\beta}$ is the known beta-factor. The DKE is less then one, if we assume $v^2\leq c^2$. For any point in space, denote by $u$ the \textit{dimensionless potential energy} (DPE)
\begin{equation}\label{u_def}
  u(\mathbf{x})=-\frac{2U(\mathbf{x})}{mc^2}.
\end{equation}
Finally, denote by $\mathcal{E}=\frac{2E}{mc^2}$ \textit{dimensionless total energy} (DTE) of the orbit.

Dividing equation (\ref{energycos}) by $\frac{1}{2}mc^2$ of the particle we obtain the unit-free equation expressing the \textit{dimensionless energy  conservation} on the trajectory
\begin{equation}\label{RE gdecompNG}
  \frac{\mathbf{v}^2}{c^2}-u(\mathbf{x})=\mathcal{E}.
\end{equation}
It follows from (\ref{2Newton}), that the acceleration $\mathbf{a}$ of a free moving particle under the given potential satisfies
\begin{equation}\label{Accel_DPE}
 \mathbf{a}=-\frac{1}{m}\nabla U =\frac{c^2}{2}\nabla u\,.
\end{equation}
Thus,
\begin{equation}\label{DPEaccel}
 u(\mathbf{x})=\frac{2}{c^2}\int_{O}^{\mathbf{x}}\mathbf{a}(\mathbf{x}(\tau))\cdot d\mathbf{x}(\tau),
\end{equation}
where the integral is over some free motion curve $\mathbf{x}(\tau)$ connecting  $O$-the zero of the potential energy to $\mathbf{x}.$
Note that DKE is defined by the velocity at the given time, while DPE is the integral of the acceleration over the trajectory.

The DPE can be expressed  by the use of the notion of \textit{escape velocity} $v_e(\mathbf{x})$,  defined as the minimum speed needed for a mass particle at $\mathbf{x}$ to ``break free" (if possible) from the  attraction defined by the potential.  More particularly, it is the velocity (speed away from the starting point) at which the sum of the particle's kinetic  and its  potential energies is equal to zero. Thus,
\begin{equation}\label{DPE_ve}
   u(\mathbf{x})=\frac{v_e^2(\mathbf{x})}{c^2},\;\; \mathbf{v}_e=-c\sqrt{u}\frac{\nabla u}{|\nabla  u|}.
\end{equation}
We will define the \textit{Schwarzschild region} the region $S$ defined by
\begin{equation}\label{Scwar}
  S=\{\mathbf{x}:\;\;v_e^2(\mathbf{x})\leq c^2 \}.
\end{equation}
Any  particle positioned in $S$ is attracted by the source of the potential. Thus, any \textit{free} motion of a mass particle is outside $S$.


We will say that a force is \textit{central} if there is a fixed point in space, called the center of the force, such that the force on a mass particle is always in the direction of the center. We choose the origin of our coordinate system to be at the center of the force and denote the position vector with respect to the origin by $\mathbf{r},$ its length by $r$.
 Thus, the potential of a central force is dependent only on the radial distance and is of the form $U(\mathbf{r})=U(r)$.

Let $\mathbf{r}(t)$ be the trajectory under a central force of a massive particle. In this case the angular momentum is conserved \cite{McCuskey}. This implies that the motion is in the plane perpendicular to the angular momentum and we may use polar coordinates for the position in this plane.  Denote by $\mathbf{J}$
the angular momentum per unit mass on the trajectory. The angular velocity on the trajectory is
\begin{equation}\label{Ang_vel}
 \frac{d\varphi}{dt}=\frac{J}{r^2}
\end{equation}
and we can  { decompose} the square of the  velocity of the particle as the sum of the squares of its  orthogonal \textit{radial} and \textit{transverse} components
 \begin{equation}\label{Theta1_dot}
 \left(\frac{d\mathbf{r}}{dt}\right)^2=\left(\frac{dr}{dt}\right)^2+\frac{J^2}{r^2}.
\end{equation}
 Substituting this into (\ref{RE gdecompNG}) we obtain the  classical \textit{dimensionless decomposed energy conservation equation}
 \begin{equation}\label{RE decompNG}
 \frac{1}{c^2}\left(\frac{dr}{dt}\right)^2 +\frac{J^2}{c^2r^2}-u(r)=\mathcal{E},
\end{equation}
where the first and second terms express the radial and transverse components of DKE respectively.

For an inverse-square law force the potential is  $U(\mathbf{r})=-\frac{k}{r}$, and by use of (\ref{DPE_ve}) and (\ref{u_def})
\begin{equation} \beta_e=\frac{v_e}{c}=\sqrt{u}=\sqrt{\frac{2k}{rmc^2}},\end{equation}
and the radius $r_s$ of the Schwarzschild region is
\begin{equation}\label{ScwCentral}
 r_s=\frac{2k}{mc^2}\;\mbox{ and }\;\beta_e^2=u=\frac{r_s}{r}\,.
\end{equation}
From (\ref{2Newton}) the acceleration of the particle  is $a= \frac{k}{mr^2}$ and from
(\ref{Scwar}) we obtain $a\leq \frac{k}{mr_s^2}=\frac{c^2}{2r_s}$. This implies that for such a motion there is an upper limit $a_m$ on the acceleration (depending on the potential) with
\begin{equation}\label{max_accel}
  a_m=\frac{c^2}{2r_s} \;\mbox{ and }\;u=\beta_e^2=\sqrt{\frac{a}{a_m}}\,.
\end{equation}

For the harmonic oscillator $\mathbf{F}=-k\mathbf{r}$ with potential  $U(\mathbf{r})=\frac{k}{2}r^2.$ Using (\ref{u_def}), this implies that $u(\mathbf{r})=\frac{\omega_0^2 r^2 }{c^2}$ where $\omega_0=\sqrt{k/m}$ is the natural frequency of the oscillator. Since the acceleration in this case is $\mathbf{a}=-\omega_0^2 \mathbf{r}$,  the expression for $u$ is now
\begin{equation}\label{DPR_ho}
  u=\frac{\omega_0^2 r^2 }{c^2}=\frac{a^2}{a_m^2}  \;\mbox{ with }\; a_m=c\omega_0.
\end{equation}
Note, that for the harmonic oscillator there is no notion of an escape velocity.

\section{Influence of the central force potential  on  space-time}

\textit{Relativistic Newtonian Dynamics} (RND) is a modification of the Newtonian dynamics by transforming it from absolute space and time to spacetime influenced by energy.

  In SR the time passage of a moving clock as observed in the laboratory frame is altered by the  Lorentz $\gamma$ factor depending only on its velocity (or DKE) with respect to the lab frame $K$. On the other hand in GR, the rate of a rest clock is altered by the gravitational time dilation factor $\tilde{\gamma}$  depending on the gravitational potential at that point.

  It is known that if a velocity is tangent to a trajectory (smooth manifold), the motion remains confined to the trajectory. Thus, time modification alone will not change the classical trajectory, since such modification changes only the magnitude of the velocity, which was tangential to the trajectory,  without changing its direction. The anomalous precession of Mercury shows however that in fact the real trajectory of the planet differs from the classical one. Thus, a relativistic model which will be able to predict the observed precession must also modify space. In SR, modification of time in a system moving uniformly comes together with a modification of \textit{one} of the directions in space, the direction of the velocity, which we will call the \textit{influenced direction}.

The observed anomalous precession of Mercury, for example, is significantly larger than the one predicted by SR which considers the influence of the kinetic energy only. Thus, in RND, we must also consider the \textit{influence of the  potential energy} on spacetime in the neighbourhood of any point in spacetime. We will compare the modified spacetime with respect to the flat spacetime of a locally inertial lab frame with zero force (potential) at its origin. To simplify the model we will require that our potential in the lab frame is independent of time.

 Rather then proposing an a priori postulate for such influence, we will be guided by the well known principles introduced by A. Einstein for the gravitational potential. Firstly, by using Einstein's Equivalence Principle, we study the influence of the potential energy in our system by considering instead the influence of  acceleration of a system in free space. This is justified, since both potential energy and acceleration produce the same effects inside these systems. Secondly, by using the notion of a \textit{comoving frame} associated to a given point in an accelerated frame as an inertial frame with the same velocity as the instantaneous velocity of  this point, and an extension of Einstein's Clock Hypothesis, we express the interconnection between  the spacetimes in the neighbourhood of two points in an accelerated system, via the Lorentz transformations between the comoving frames at these points.

 Einstein's Equivalence Principle (1907) states \cite{Einstein07} `` we [...] assume the complete physical equivalence of a gravitational field and a corresponding acceleration of the reference system ". To be able to use this principle effectively, we must specify the meaning of ``Corresponding Accelerated System". Obviously, such system will depend on $\mathbf{r}_0$, the position in space outside the Schwarzschild region $S$. Our accelerated system will be described by the collection of spacetime frames $K_{r_0}(u)$ parameterized by $u$ (DPE defined by (\ref{u_def})) attached to an accelerated observer.

 Define the \textit{escape trajectory} as the (de)accelerated trajectory starting at $\mathbf{r}_0$ corresponding to DPE $u_0$ of a mass particle with escape velocity $\mathbf{v}_e$, defined by (\ref{DPE_ve}), and progressing freely with decreasing potential to the ultimate point $O$ with zero potential.  We propose that the Corresponding Accelerated System in the   Equivalence Principle is $K_{r_0}(u)$ - the collection of Frenet frames \cite{FS3} defined by this escape trajectory starting with $K'=K_{r_0}(u_0)$ and ending with $K=K_{r_0}(0)$ - the inertial lab frame.

  On the escape trajectory the total energy in $K_{r_0}(u_0)$ is $\mathcal{E}=0$  by the definition of $\mathbf{v}_e$ and remain, by energy conservation,  zero in all $K_{r_0}(u)$.  Thus, the velocity of the mass particle at each point of this trajectory is equal to the escape velocity at this point. For a central force the direction of this velocity is constant and equal to the radial direction $\mathbf{r}_0/r_0$, hence from (\ref{DPE_ve})
  \begin{equation} \frac{dv_e}{dt}=\frac{dv_e}{du}(\nabla u\cdot\frac{d\mathbf{r}}{dt})= -\frac{c}{2\sqrt{u}}\nabla u (-c\sqrt{u})=\frac{c^2}{2}\nabla u,\end{equation}
 which is by (\ref{Accel_DPE}) the acceleration under the given potential.
The escape trajectory provides a connecting road between the frame $K'$  at $\mathbf{r}_0$ with potential in it and the inertial lab frame $K$ at $O$ with zero potential.

Note that the frames $K_{r_0}(u)$ are inertial frames in the presence of the gravitational field, but they are accelerated with respect to $K$ if  the gravitational potential is ignored.
Finally, by using an extension of Einstein's Clock Hypothesis, introduced in \cite{FS2,FS3} and \cite{Mash1,Mash2}, we can interconnect the spacetime transformations between  the two accelerated frames $K'$ and $K$ by use of the Lorentz transformations between the comoving frames associated to their origins.


 For a central force potential, the frame $K'$  moves with escape velocity $\mathbf{v}_e $ in the radial direction with respect to $K$. If we choose the first spacial coordinate in the radial direction, and denote by $\beta_e^2=v_e^2/c^2$, the spacetime transformation  (Lorentz transformations) from $K'$ to our reference frame
 $K$ is
 \begin{eqnarray}\label{FFFbasis}
    \nonumber ct &=& \tilde{\gamma}(ct'+\beta_ex'_1),\;\;\; x_2 = x'_2,\\
    x_1 &=& \tilde{\gamma}(\beta_e t'+x'_1),\;\;\; \;x_3 = x'_3,
    \end{eqnarray}
 where
 \begin{equation}\label{gammatilde_def}
 \tilde{\gamma}=\frac{1}{\sqrt{1-v_e^2/c^2}}=\frac{1}{\sqrt{1-u_0}}
\end{equation}
 is the known gravitational \textit{time dilation factor}.

The time dilation factor $\tilde{\gamma}$ at any given point $\mathbf{r}$ depends on the DPE, defined by the potential energy with the boundary condition, which is a non-local property. We can sometime define it explicitly from a measurable value on the trajectory, the acceleration, but it will differ for different types of force dependence. For example for an inverse-square law force using (\ref{max_accel}) this factor is
\begin{equation}\label{TimeDilisl}
  \tilde{\gamma}=\frac{1}{\sqrt{1-\sqrt{\frac{a}{a_m}}}}
\end{equation}
which differs significantly from the time dilation factor in all models of maximal acceleration \cite{FG10}-\cite{F_Ann}.

To define the influence of the potential energy for a harmonic oscillator the escape trajectory is replaced by a trajectory of a particle with 0-velocity  at $\mathbf{r}_0$ moving freely  to the origin of the oscillator $O$. Using (\ref{DPR_ho}), the velocity $\mathbf{v}_0$ at $O$ on this trajectory will be in the radial direction with magnitude $v_0=\omega_0 r$. For this trajectory, the above arguments nevertheless remain valid if one interchanges $v_e$ with $v_0$. Thus, the time dilation factor in this case, using (\ref{DPR_ho}), is
\begin{equation}\label{TimeDilHO}
  \tilde{\gamma}=\frac{1}{\sqrt{1-\frac{a^2}{a_m^2}}},
\end{equation}
which is the time dilation factor in all models of maximal acceleration.

 From the transformation formulas (\ref{FFFbasis}) we get
 \begin{equation} dt=\tilde{\gamma}dt',\;dx_1=\tilde{\gamma}dx'_1,\;dx_2=dx'_2,\;dx_3=dx'_3\,\end{equation}
 implying that the 3D velocity transformations between these systems is
 \begin{equation}\label{valoc_transf}
   (v_1,v_2,v_3) =(v'_1,\tilde{\gamma}^{-1}v'_2,\tilde{\gamma}^{-1}v'_3).
 \end{equation}
Thus the influence of acceleration or potential energy at $x_0$ on any velocity is expressed by multiplication of the component of this velocity transverse to  $\nabla U({x}_0)$ by $\tilde{\gamma}^{-1},$ where the time dilation factor $\tilde{\gamma}$ is defined by (\ref{gammatilde_def}).

 \section{RND equations of motion under a central force}

For a central force, the direction influenced by the potential energy is the radial direction implying that the radial velocities are not affected by this influence, while the transverse ones should be  multiplied by $\tilde{\gamma}^{-1}=\sqrt{1-u(r)}$. This implies, that in our lab frame $K$ the decomposition of the square of the  velocity of the particle as the sum of the squares of its  orthogonal radial and transverse components, given by (\ref{Theta1_dot}), should be modified to
\begin{equation}
 \left(\frac{d\mathbf{r}}{dt}\right)^2=\left(\frac{dr}{dt}\right)^2+\frac{J^2}{r^2}(1-u(r)).
\end{equation}
Thus, considering the influence of the potential energy, the dimensionless energy conservation equation (\ref{RE decompNG}) becomes
\begin{equation}\label{RND decompNG}
 \frac{1}{c^2}\left(\frac{dr}{dt}\right)^2 +\frac{J^2}{c^2r^2}(1-u(r))-u(r)=\mathcal{E}.
\end{equation}
This equation together with equation (\ref{Ang_vel}) form a first order system of differential equations with respect to $r(t),\varphi(t)$. They are the \textit{RND equations of motion under a central force}.

For an inverse-square law force, using known methods (see for example \cite{McCuskey} and \cite{FS}) we can obtain the trajectory of the motion by solving equation (\ref{RND decompNG}) for the function $u(\varphi)$. If we denote $u'=\frac{du}{d\varphi}$, then from (\ref{ScwCentral}) and (\ref{Ang_vel}) we obtain $\frac{dr}{dt}=-\frac{J}{r_s}u'$. Substituting this in the above equation give
\begin{equation}\frac{J^2}{c^2r_s^2}(u')^2 +\frac{J^2u^2}{c^2r_s^2}(1-u)-u=\mathcal{E}.\end{equation}
Multiplying this equation by $2\mu$ where
\begin{equation}\label{mu_def}
  \mu=\frac{c^2r_s^2}{2J^2},
\end{equation}
we obtain
\begin{equation}\label{uprime_eqn}
  (u')^2 = u^3-u^2 +2\mu u +\tilde{e}
\end{equation}
for some constant $\tilde{e}$. As it was shown in \cite{FS} this equation defines a precessing trajectory with precession $3\pi\mu$ radians per revolution, with $\mu$ as the average value of $u$ on this trajectory and is also equal $\mu=\frac{r_s}{L},$
where $L$ is the semi-latus rectum of the  orbit.

The Schwarzschild radius of the Sun is $r_s=2953.25 m$, for Mercury $L=5.546\cdot10^{10}m$  implying that   $\mu= 5.32497\cdot10^{-8}$. Thus, our model predicts a $5.01866\cdot10^{-7}$ radians per revolution precession of the perihelion of Mercury, which is \textit{exactly} the currently observed one.


 \section{RND of Hydrogen like atom}

 Consider a system of two particles, a proton with mass $m_p=1.7\cdot10^{-27} kg$ and an electron with mass $m_e=9\cdot10^{-31}kg$.
Denote by $\mathbf{r}_p, \mathbf{r}_e$ the positions of the proton and the electron, respectively and the relative position of the electron with respect to the proton by $\mathbf{r}=(\mathbf{r}_e-\mathbf{r}_p)$. We will ignore the interaction of the particles with the fields and restrict ourselves only to the Coulomb force. The force of the field of the proton acting on the electron is thus $\mathbf{F_1}=-k\mathbf{r}/r^3$, with $k=2.3\cdot10^{-28}Nm^2$, while the electric force of the electron acting on the proton is $\mathbf{F_2}=k\mathbf{r}/r^3=-\mathbf{F_1}.$

The center of mass of this two-particle system is positioned at or close to the proton, and the electron moves around the more or less stationary proton.
The motion of the electron is a motion under a inverse-square law force of the Coulomb field of the proton with potential
\begin{equation}\label{HApot}
  U(\mathbf{r})= -\frac{k}{r}.
\end{equation}

The dimensionless potential energy $u$ and the Schwarzschild radius $r_s$ defined by (\ref{ScwCentral}) are
\begin{equation}\label{SchwarzHy}
  u(\mathbf{r})=\frac{2k}{m_ec^2r}=\frac{r_h}{r},\;\;\; r_s=\frac{2k}{mc^2}=5.7\cdot 10^{-15}m
\end{equation}
Since the typical distances between the proton and the electron are of the order $r_0=0.5A=0.5\cdot10^{-10}m$, we can estimate the value of $\mu$ defined by (\ref{mu_def}) as the value of $u$ at this average distance
 \begin{equation}\mu\approx\frac{r_s}{r_0}=\frac{5.7\cdot 10^{-15}}{0.5\times10^{-10}}\approx10^{-4}\,.\end{equation}
Thus, from the previous Section,
the precession of the trajectory is approximately
 \begin{equation}\triangle\varphi=3\pi\mu\approx10^{-3} rad/rev\end{equation}

 The period of the orbit by \cite{McCuskey} is  {$T=2\pi r_0^{3/2}/\sqrt{k/m_e}=0.55\cdot10^{-14}s$}, implying that the number of revolutions per second is  { $2\cdot10^{14}$}. Thus, the precession per second  is  {$2\cdot10^{11} rad$}
 and the precession per nanosecond is  {$200$} radian. Since a typical measurement of distance takes nanoseconds, the electron will cover a full area between the two radial distances $r_1$ and $r_2$ depending on the initial conditions, see Figure 1. {Thus, the RND model reveals that the dynamics of the electron in the Hydrogen like atom is chaotic. This explains the quantum mechanics probabilistic model its description.}
  \begin{figure}
  \centering
\scalebox{0.3}{\includegraphics{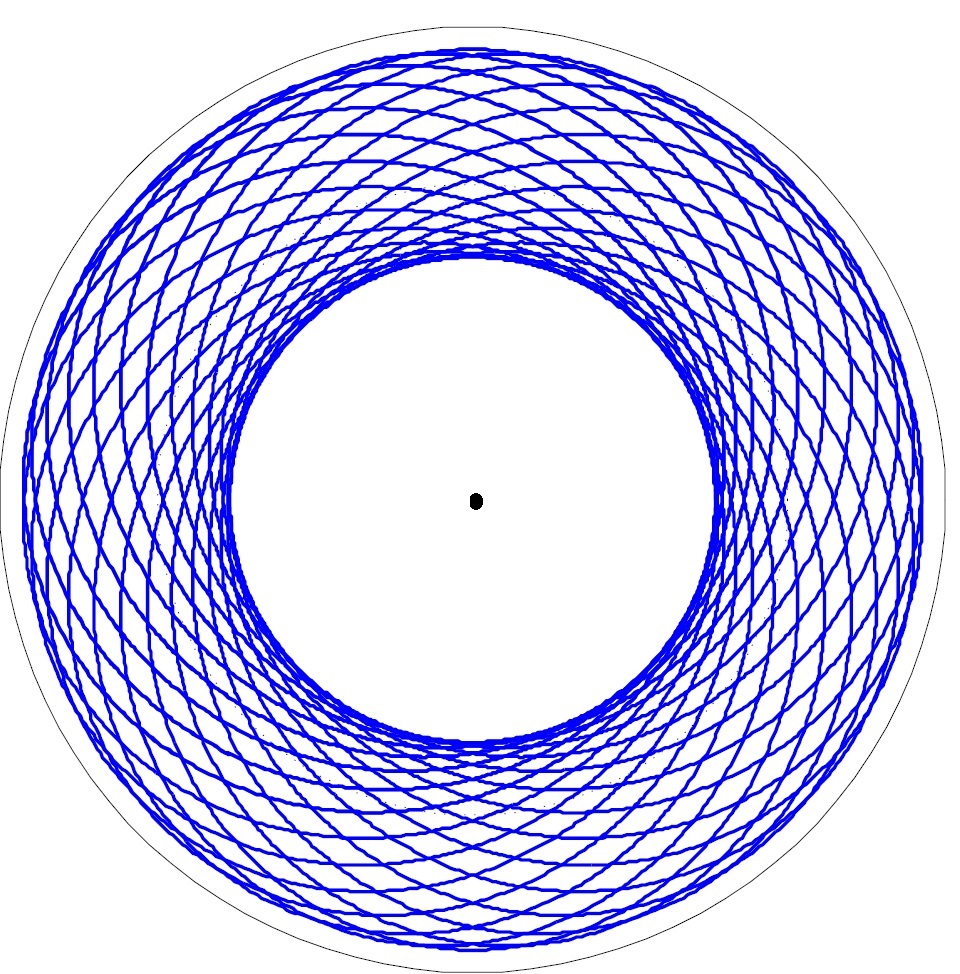}}
  \caption{Electron and proton trajectories in ERD Hydrogen like atom}\label{MaxAccelExp}
\end{figure}
In general, our solution  is not a closed path, but for certain discrete initial values, the solutions will be periodic.

\section{Discussion}

Based on Einstein's Equivalence Principle for the gravitational potential and an extension of his Clock Hypothesis, we proposed a description of the influence of the potential energy on spacetime by (\ref{FFFbasis}) and (\ref{gammatilde_def}). For the harmonic oscillator potential, the dependence of the time dilation factor on acceleration is given by (\ref{TimeDilHO}) and is similar to the factors used in other models \cite{FG10}-\cite{Torrome} considering the influence of acceleration (the gradient of the potential) on spacetime. But for other types of potentials it differs significantly. For example for inverse-square law force potential this dilation factor is given by (\ref{TimeDilisl}).

For a conservative central force potential we obtained a relativistic extension of Newton's dynamics (RND) by incorporating the influence of the potential energy on spacetime into the relative energy conservation equation (\ref{RND decompNG}) and use of the angular momentum conservation equation (\ref{Ang_vel}). Applied to motion under inverse-square law force this dynamics, as in \cite{FS}, predicts precessing trajectories and predicts the exact precession of Mercury. For a Hydrogen-like atom it explain the probabilistic behaviour of the orbit.

We plan to extend the model to motion under a general  force. For this model we need to derive an expression for a Doppler type shift caused by the potential. Such shift was observed recently in experiments \cite{F14} and \cite{F16_EPL} testing Einstein's Clock Hypothesis by use of rotating M\"{o}ssbauer absorber and synchrotron radiation. By use of RND we will try explain the experimental results of \cite{Potzel} of testing Clock hypothesis by use of M\"{o}ssbauer spectroscopy.

\acknowledgments I wish to thank Prof. J. M. Steiner for the  discussions which helped to formulate and clarify the results of this paper and Prof. Lawrence Paul Horwitz for constructive comments.

\end{document}